\documentstyle[prb,multicol,aps,epsfig,epsf,amstex]{revtex}

\begin{document}
\draft
\title{Fermionic random transverse-field Ising spin chain}  
\author{A.L. Chudnovskiy $^\ast$}  
\address{Institut f\"ur Theoretische Physik, Universit\"at W\"urzburg, 
D--97074 W\"urzburg, F.R.Germany}  
\date{\today}
\maketitle


\begin{abstract}
The interplay of spin and charge fluctuations in the random 
transverse-field Ising spin chain on the fermionic space is  
investigated. The finite chemical potential, which controls the 
charge fluctuations, leads to the appearance of the quantum 
critical region in the phase diagram where the magnetic correlations 
are quenched by nonmagnetic sites.  Regions of nonmonotonous 
temperature dependence of spin-spin correlation length 
appear at nonzero $\mu$. The results on the one-fermion density 
of states of the model are discussed. \\ 
\end{abstract}

\section{Introduction}

The random transverse-field Ising spin chain (RTFISC) is the 
simplest nontrivial model exhibiting a quantum phase transition 
driven by quenched random interactions.
  The investigations of this 
model allowed to obtain many results which are of general 
importance for the quantum phase transitions in systems with 
random interactions  \cite{{f1},{f2},{shan},{rieger}}. 

The extension of the RTFISC model on the fermionic space brings 
two  aspects. From one hand, the original spin chain becomes a 
system of fermions with (statistical) charge fluctuations.  That allows 
us to investigate the interplay of the charge and spin 
fluctuations at quantum phase transition. 

Quantum phase transitions in fermionic systems with quenched 
disorder affect significantly the properties in  both spin and charge 
sectors, which has been emphasized in a group of theoretical works 
\cite{s1,s2,sen-g,s3,sro,ro1} and suggested by experimental results 
for heavy fermion systems \cite{{maple},{loehne}}.  It will be shown 
in this paper that the statististical charge fluctuations affect the 
magnetic phase diagram of the chain, creating a quantum critical 
region close to the critical point. In that region the role of the charge 
fluctuations is analogous to the role of the thermal fluctuations at small 
finite temperature \cite{s3,sro}. 
The fermionic description also represents a first step toward 
the introduction of electron hopping, which would allow us 
to investigate transport properties and a spatial structure of the 
electronic states \cite{falko}. 

On the other hand, the extension of a spin model on the 
fermionic space provides the opportunity of a natural description 
of dilution of the magnetic model with nonmagnetic (empty and 
doubly occupied) sites. Working in a great canonical ensemble at 
a fixed chemical potential, one describes the  annealed dilution. 

Phase transitions in diluted magnetic systems have been a subject 
of numerous investigations (for a review, see Refs. 
\cite{{stinch},{chakr}}). 
In pure systems the annealed dilution leads to aggregation effects, 
whereas the separation of thermal and percolating contributions to 
correlation functions is proper to the quenched dilution \cite{stinch}.

The RTFISC represents a case of the annealed site dilution in a 
system  with quenched magnetic randomness.  Due to the 
interplay between the quenched random magnetic interactions 
and the annealed dilution, the separation of magnetic and 
percolating contributions to the inverse spin-spin correlation 
length  combines with the dependence of the dilution probability 
on the strength of magnetic fluctuations. 

In this paper we  analyze the random transverse-field Ising spin 
chain on the fermionic space  (later referred to as fermionic chain), 
concentrating especially on the effects of a finite chemical 
potential $\mu$ on the spin-spin correlation function and on the 
one-fermion density of states. The presented analysis is based on 
the real-space renormalization group (RG) method applied by 
D. Fisher in the paper \cite{f1}, further referred to as I.  

It was shown in the work I, that the near-critical properties of 
the chain on the spin space (later referred to as spin chain) are 
described by the universal scaling distributions of the renormalized 
bonds and transverse fields. 
Our considerations show that the behavior of the fermionic chain 
is, however, nonuniversal and depends on the {\it original} distribution 
of transverse fields. The effect of bonds on the properties of the 
fermionic chain is described by the universal scaling distributions 
of the Ref. I.

The duality 
between the bonds and transverse fields, which is an important 
feature of the spin chain, is destroyed by the 
charge  fluctuations in the fermionic chain. 
An interesting feature of the fermionic chain is the appearance 
of  regions in the phase diagram at finite $\mu$, where the 
spin-spin correlation length changes nonmonotonously with temperature. 
In what follows, we provide an explanation for such quasireentrant 
behavior.

\section{General consideration}

In the great canonical ensemble, the Hamiltonian of the model under 
consideration reads 
\begin{equation}
\hat{H}=\sum_i 
\left\{-J_i\hat{S}_i^z\hat{S}_{i+1}^z 
-h_i\hat{S}_i^x-\mu\hat{n}_i\right\},
\end{equation} 
where $\hat{S}_i^\nu\equiv\hat{a}_{i,\alpha}^+ 
\sigma^\nu_{\alpha,\beta} \hat{a}_{i,\beta}$ is the operator of a  
$\nu$-component ($\nu=x,z$) of the spin of the fermion at the site $i$, 
$\hat{n}_i=\hat{a}^+_{i,\alpha}\hat{a}_{i,\alpha}$ is the 
particle number operator at the site $i$, $\hat{a}_{i,\alpha}^+$ and  
$\hat{a}_{i,\alpha}$ are the fermion operators representing a fermion 
on the site $i$ with a $z$-component of the spin $\alpha=\pm 1$, 
and $\sigma^{\nu}_{\alpha,\beta}$ denotes the element of 
the  Pauli matrix. 

Since there is no explicit charge dynamics (hopping) in the model, 
the statistical charge fluctuations (the nonmagnetic sites with occupation 
number 0 or 2)   are determined by the  distribution of the energy levels 
in the system. The distribution of the energy levels in turn follows from 
the distributions of random 
bonds and transverse magnetic fields. The strength of the charge 
fluctuations (the number of nonmagnetic sites) is controlled by the 
chemical potential and temperature. 

At zero chemical potential and temperature, all sites in the chain are 
magnetic and the phase diagram of the fermionic chain is equivalent 
to that of the spin chain. The fermionic chain undergoes a quantum 
phase transition when the distributions of transverse magnetic fields 
$h$ and ferromagnetic couplings $J$ 
satisfy $\delta = 0$ with $\delta$ defined in I as 
$\delta\equiv \frac{\overline{\ln h}-\overline{ln J}}{var(\ln h) 
+var(\ln J)}$.

At zero temperature, a  nonmagnetic site breaks magnetic correlations 
between the spins on different sides of it. Due to the random space 
positions of the nonmagnetic sites, the whole chain is divided into 
magnetic fragments of random lengths. Therefore, the nonmagnetic sites 
introduce edge effects to spin-spin correlations in magnetic fragments. 

Generally, the probability of a given configuration of nonmagnetic sites 
in the chain is a functional of the whole configuration. The situation 
simplifies, if the nonmagnetic sites occur very rarely 
in the chain, i.e. the distance between the neighbor nonmagnetic sites is large.  
Then, neglecting the edge effects of the nonmagnetic sites on the spin-spin 
corelations,  
one can approximately describe the distribution of nonmagnetic sites in 
terms of the probability of a single nonmagnetic site in the otherwise 
magnetic chain. We denote that probability as $W(\mu)$. 

Given $W$, an important 
relation between the spin-spin correlation functions on the spin space 
and on the fermionic space can be derived.  Since a nonmagnetic site 
breaks completely the correlation between spins on different sides of it, 
the spin-spin correlation function on the fermionic space 
$\langle S_i^z S_{i+x}^z\rangle$ 
equals the correlation function on the spin space, if all the sites between 
$i$ and $i+x$ are magnetic, and  is zero otherwise. For the mean 
spin-spin correlation functions, we obtain 
\begin{equation}
C_f(x)=(1-W)^x C_s(x), 
\end{equation}\label{cf} 
where $C_f(x)$ is the correlation function on the fermionic space, 
$C_s(x)$ is the correlation function on the spin space, and $(1-W)^x$ 
expresses the probability that all the sites between $i$ and $i+x$ are 
magnetic. Note that the factorization of the correlations in  
Eq. (\ref{cf}), 
as well as the expression $(1-W)^x$ for the probability of a nonmagnetic 
fragment with length larger than $x$, is valid only in the limit of rare 
nonmagnetic sites. Using expression (\ref{cf}), we obtain the leading 
contribution to the spin-spin correlation function at small $W$.   

At $W\ll 1$ we can write, approximately, 
\begin{equation} C_f(x)=e^{-Wx} C_s(x),
\label{cfexp} \end{equation}
then the influence of the nonmagnetic sites at finite $\mu$ reduces to 
an additive contribution to the inverse correlation length. The separation 
of the inverse correlation length on the magnetic and nonmagnetic 
contribution is typical of one-dimensional systems with quenched 
dilution \cite{stinch}.

Close to the critical point, the magnetic spin-spin correlation length 
diverges. Then, at finite $\mu$, the correlation length of the 
fermionic chain is determined by the value $1/W(\mu)$, which 
is the average distance between two neighbor nonmagnetic sites. 
By analogy with the effect of small temperature on the quantum phase 
transition \cite{{s3},{sro}}, we call the region, where $W(\mu)$ 
dominates over the magnetic inverse correlation length, the quantum 
critical region.  The relation between $W(\mu)$  and the inverse 
correlation length of the spin chain determines the size of the quantum 
critical  region.

In the quantum disordered phase the spin-spin correlation length 
can be represented in the form
\begin{equation}
\xi=\delta^{-2}\frac{1}{1+W/\delta^{2}}, 
\label{xws}
\end{equation}
where $\delta^2$ is the inverse correlation length on the spin space found 
in I. 
From  Eq. (\ref{xws}), one identifies formally the quantum 
fluctuations -- percolation 
crossover exponent $\phi=2$ (Ref. \cite{stinch}). At the same time, the 
probability $W$ is a function 
of $\delta$, $W=W(\mu,\delta)$, and the relation between the 
chemical potential $\mu$ and the distance from the critical point $\delta$ 
at the crossover cannot be expressed in a power-low form.

In order to obtain the probability $W(\mu)$, we follow a given 
site during the applied in the paper I RG-process and find its 
ground state. To this end, consider an ensemble of chains with the 
transverse field at a given site $P$ fixed to the value $h_p$. The 
probability of this ensemble is $\rho(h_p)$, where $\rho(h)$ is 
the {\it original} distribution of the transverse fields. The next to 
the site $P$ bonds, $J_L$ and $J_R$, are random, and their distributions 
evolve in the RG process according to the RG-equations.

The main assumption of the RG procedure in I 
is that of broad near-critical distributions of renormalized 
bonds $J$ and fields $h$. It follows from this assumption that the 
strongest transverse fields, $h_i=\Omega=max\{J_j,h_j\}$, and  the strongest  
bonds, $J_i=\Omega$, determine 
completely the ground states of the associated local parts of the chain,  
the couplings to the rest of the system being treated perturbatively. 
For the renormalization scale $\Omega\gg\mu$, the 
sites with couplings of the strength $\Omega$ are singly occupied 
and the RG transformation is the same as in I. The spins at 
the sites with $h_i$ in the infinitesimal interval $d\Omega$ below 
$\Omega$ are decimated  to form effective bonds of the strength 
$\tilde{J_i}=J_{i-1}J_i/\Omega$, and the spins at the neighbor sites 
with the bond $J_i$ in the interval $[\Omega-d\Omega,\Omega]$ form spin 
clusters with effective transverse fields $\tilde{h}_i=h_ih_{i+1}/\Omega$.  
Then the RG scale $\Omega$ is lowered infinitesimally by $d\Omega$.   

At  $\mu\ll\Omega_I$, where $\Omega_I$ 
is the initial RG-scale, the typical length of magnetic 
fragments is large, and, neglecting the edge effects of the nonmagnetic sites, 
we assume the distributions of couplings in a  magnetic fragment to be 
described by the RG-solutions of the 
paper I for the distributions of couplings in the infinite chain.

The ground state of the site $P$ becomes defined in the 
moment when it is decimated in the RG 
process.  The magnetic or nonmagnetic character of the ground state 
depends on the energy balance between the lowest 
magnetic and nonmagnetic states.

The site $P$ can assume a nonmagnetic ground state in two 
ways. \\ 
{\bf (i)} At a scale $\Omega>h_p$, the site is coupled (for example, 
from the right) with the strongest bond $J_R=\Omega$ to a 
spin cluster. In this case, the small contribution of the other 
adjacent bond $J_L$ to the energy of the magnetic state 
should be taken into account.  The energy of the magnetic 
state is $E_m=-\Omega-J_{L}-\mu$ and the energy of the 
nonmagnetic state is $E_{nm}=-2\mu$. The site becomes 
nonmagnetic if $\mu-\Omega-J_{R,L}>0$. \\ 
{\bf (ii)} The site is not decimated up to the scale $\Omega=h_p$. 
Then the transverse field $h_p$ becomes the strongest coupling. 
The energy of the magnetic state is $E_m=-\Omega-\mu$ and the 
energy of the nonmagnetic (doubly occupied) state is $E_{nm}=-2\mu$. 
Therefore, the ground state is nonmagnetic if $\mu-\Omega>0$. 

The ground state of a once decimated site does not change in the 
following RG transformations. 

According to the assumption of rare nonmagnetic sites, we neglect 
the pairs of nonmagnetic sites and larger nonmagnetic 
clusters. 

The site $P$ remains undecimated to the scale $\Omega$ if the 
{\it original} transverse field $h_p$ is smaller than $\Omega$ and 
none of the renormalized adjacent bonds $\tilde{J}_{R,L}$ takes 
the strongest value from the beginning of the RG procedure up to 
$\Omega$. 

On the logarithmic scale $\Gamma=\ln(\Omega_I/\Omega)$, we 
define the "survival" probability $S(\zeta, \Gamma)$. $S(\zeta, \Gamma)$ 
is the probability  that the effective bond $\tilde{J}_R$, adjacent 
from the right to the site  $P$,  never takes the strongest value (survives) 
up to the scale $\Omega>h_p$ and equals $\tilde{J}_R= \Omega\exp(-\zeta)$ 
at the scale $\Omega$. The function  $S(\zeta, \Gamma)$ 
satisfies the following RG-equation: 
\begin{equation}
\frac{\partial S(\zeta, \Gamma)}{\partial \Gamma} = \frac{\partial 
S(\zeta, \Gamma)}{\partial \zeta} - R_0(\Gamma) S(\zeta, \Gamma) + 
R_0(\Gamma) S \underset{\zeta}{\otimes} P,
\label{S}\end{equation}
where $R_0(\Gamma)=2\delta/(1-e^{-2\delta(\Gamma+C_0)})$ is the 
probability of the strongest transverse magnetic field at the logarithmic 
RG scale $\Gamma=\ln(\Omega_I/\Omega)$, $P(\zeta, \Gamma)$ is 
the probability density of the bond of the logarithmic strength $\zeta=\ln(\Omega/J)$, 
and $C_0$ is a RG-irrelevant constant that can be 
obtained from the original distributions of couplings $\pi(J)$ 
and $\rho(h)$. 
The functions $R_0(\Gamma)$ and $P(\zeta, \Gamma)$ were found in the work I. 
The explicit view of $P(\zeta, 
\Gamma)$ is $P(\zeta, \Gamma)=u(\Gamma) e^{-\zeta u(\Gamma)}$ 
with 
$u(\Gamma)=2\delta/(e^{2\delta(\Gamma+ C_0)}-1)$. 

The first two terms in Eq. (\ref{S}) describe the reduction of 
the probability due to the decimation of the bonds of the strength 
$\Omega$ and the overall rescaling of the energies in the RG 
transformation. The last two terms describe, respectively, 
the reduction of the bonds and 
the creation of effective bonds of the logarithmic strength $\zeta$ 
by the decimation of the right neighbor site 
due to the strong transverse field. 
The last term is 
proportional to the convolution of the distributions of the bonds 
adjacent to the right neighbor site, 
$S\underset{\zeta}{\otimes} P\equiv \int_0^\zeta d\zeta' 
S(\zeta',\Gamma) P(\zeta-\zeta', \Gamma)$.  Equation (\ref{S}) 
looks dual to Eq. (5.3) of paper I, which describes the survival 
probability of an end-point spin cluster. 

The special solution of  Eq. (\ref{S}), satisfying the initial 
normalization $\int_0^\infty d\zeta S(\zeta, \Gamma=0)=1$, reads: 
\begin{equation}
S(\zeta, \Gamma)=\delta e^{-C_0\delta} 
\frac{\sinh(C_0\delta)}{\sinh^2[(C_0+\Gamma)\delta]} e^{-\zeta 
u(\Gamma)}.
\end{equation}

The equation for the probability $S(\zeta, \Gamma)$ and its 
solution are valid for the small values of the chemical potential 
$\mu\ll\Omega_I$. In this case the main contribution to the 
probability $S(\zeta, \Gamma)$ comes from the late stages of 
the RG process, where the role of the RG-irrelevant nonuniversal 
parts of the distributions of effective bonds and transverse fields is 
minor.  

The probability of the nonmagnetic ground state of the site with 
a fixed transverse field $h_p$, $W_h(\mu)$, is given now as the sum of 
probabilities for two possible cases, $W_h=w_J+w_h$.  $w_J$ 
is the probability that the nonmagnetic site appears when one of 
the adjacent bonds, say $J_R$, takes the strongest value and the 
condition of the nonmagnetic ground state $\Omega+ J_L<\mu$ 
is fulfilled. $w_h$ is the probability that none of the adjacent 
couplings becomes the strongest up to the scale $\Omega=h_p$ 
and then, at $h_p<\mu$, the site assumes the nonmagnetic ground state. 

Both probabilities, $w_J$ and $w_h$, can be expressed through 
the "survival" probability $S(\zeta, \Gamma)$ in the following manner:
\begin{equation}
w_J=2\int_{\Gamma_\mu}^{\Gamma_h}d\Gamma 
\int_{-\ln\left(\mu/\Omega_I-e^{-\Gamma}\right)}^\infty d\xi 
S(0,\Gamma) S(\xi,\Gamma),
\label{wj}\end{equation}
\begin{equation}
w_h=\left[\int_0^\infty d\zeta S(\zeta,\Gamma_h)\right]^2 
\vartheta(\mu-h)= \frac{e^{2\Gamma_h\delta} 
\sinh^2(C_0\delta)}{\sinh^2[(C_0+\Gamma_h)\delta]}  \vartheta(\mu-h).
\label{wh}\end{equation}
The prefactor 2 in Eq. (\ref{wj}) accounts for the fact that any 
of the two adjacent bonds can assume the strongest value. 

The probability $W_h$ is universal in the sense, that it depends 
on the scaling distribution of renormalized bonds and the details 
of the original distributions are irrelevant.

The whole probability $W(\mu)$ obtains as an integral 
$W(\mu)=\int_0^\mu dh \rho(h) W_h(\mu)$, which is the 
average of the probability $W_h(\mu)$ over the 
{\it original} distribution of transverse fields $\rho(h)$. 
The probability of the nonmagnetic site is therefore nonuniversal 
and depends of the details of the distribution of transverse fields 
$\rho(h)$. However, it contains a universal factor $W_h(\mu)$, 
which can be separated from the nonuniversal one.

\section{Spin-spin correlation function at zero temperature.}

Below we analyze the behavior of the function $W_h(\mu)$ in 
different regions of the phase diagram and the behavior of the 
spin-spin correlation length $\xi_f$ for two distributions of original 
transverse fields: 

a) the rectangular distribution of width $b$, $\rho(h)=(1/b) \vartheta(b-h)$ 
and 

b) the exponential distribution $\rho(h)=(1/b) \exp(-h/b)$. 

As it was mentioned above, the inverse correlation length is 
the sum of the magnetic and the nonmagnetic contributions. 
The magnetic contribution is equal to that of the model on the 
spin space and the nonmagnetic one equals $W$. We consider 
different phases separately. 

\subsection{Deep in the quantum disordered phase, $\delta>0$}

In this region one can obtain the leading contribution 
for the probability density of the nonmagnetic ground state 
at $h<\mu$ analytically,  neglecting the influence of the weak 
neighbor bond in the calculation of $w_J$. $W_h$ obtains in 
the form: 
\begin{equation}
W_h\approx \sinh^2(C_0\delta) 
\frac{e^{2\Gamma_\mu\delta}}{\sinh^2[(\Gamma_\mu+C_0)\delta]}
\approx 4e^{-2C_0\delta}\sinh^2(C_0\delta) 
\left(1-2\tilde{\mu}^{2\delta}e^{-2C_0\delta}\right),
\label{Wh1}\end{equation}
where $\Gamma_\mu\equiv -\ln(\tilde{\mu})$, 
$\tilde{\mu}\equiv\mu/\Omega_I\ll 1$ is a reduced chemical 
potential. 

The value $\tilde{\mu}^{2\delta}$ orders the expansion of 
$W_h$ at small $\tilde{\mu}$. The $\mu$-independent part of 
$W_h$ is the probability that the site remains undecimated up to 
the scale $\Omega=h$ in the chain on the spin space.  The leading 
term of the probability $W_h$ is $h$-independent, which leads 
to the factorization of the nonuniversal part of the probability 
of the nonmagnetic state $W(\mu)$,                   
\begin{equation} W(\mu)=W_h \int_0^\mu \rho(h) dh. 
\end{equation}

In the case of the rectangular distribution (a) with the width 
$b>\mu$, the 
spin-spin correlation length obtains 
\begin{equation}
\xi_f\approx \left[(\mu/b) 
\frac{e^{2\Gamma_\mu\delta} 
\sinh^2(C_0\delta)}{\sinh^2[(\Gamma_\mu+C_0)\delta]} 
+\delta^2\right]^{-1}\label{x1a} \end{equation} 
The first term in the square brackets is the nonmagnetic 
contribution and the second one is that obtained in the paper I 
contribution of the magnetic quantum fluctuations.

The correlation length for the distribution (b) obtains from 
Eq. (\ref{x1a}) by the substitution of the prefactor  
$(\mu/b)\rightarrow (1-e^{-\mu/b})$. The correlation length 
decreases with increase of $\delta$, i.e., deeper into the quantum 
disordered phase. The account for the weak couplings in calculation 
of $w_J$ does not change this dependence.

\subsection{Ordered phase, $\delta<0$}

In the ordered phase the spontaneous magnetization is suppressed 
at finite chemical potential by the presence of the nonmagnetic sites. 
The correlation length equals $\xi_f=1/W(\mu)$. The expressions 
for $W_h$ and $W$,  obtained for the disordered phase, are valid here  
if considered at $\delta<0$.  The leading behavior of the function 
$W_h(\mu)$, and hence of the correlation length at small $\tilde{\mu}$ 
is, however, different. At $\tilde{\mu}\ll 1$ we obtain 
\begin{equation} W_h\approx 
4\tilde{\mu}^{4|\delta|}e^{-2C_0\delta}.  \end{equation}
The function $W(\mu)$ differs from $W_h(\mu)$ by the 
prefactor $\int_0^\mu \rho(h) dh$, which was calculated for the 
two distributions $\rho(h)$ in the previous subsection. Now 
$W_h$, and hence $W$, decreases with $|\delta|$, i.e., the correlation 
length $1/W$ increases deeper into the ordered phase.

\subsection{Close to the critical point. Quantum critical region.}

Close to the critical point, for $\tilde{\mu}\gg e^{-1/\delta}$, 
the expansion of $W_h(\mu)$ in powers of $\mu^{2\delta}$ is 
not valid anymore, and one should make a double expansion in 
$1/|\ln\tilde{\mu}|$ and $\delta$.  The leading term of $W_h(\mu)$  
close to the critical point reads 
\begin{equation}
W_h\propto \frac{C_0^2}{(C_0-\ln\tilde{\mu})^2}.
\end{equation}\label{Wqc}
The leading term of $W_h$, and hence of the nonmagnetic 
contribution to the inverse correlation length, is $\delta$-independent. 
It is determined by the chemical potential alone, and this regime can 
be associated with the quantum critical (QC) regime of the fermionic 
chain. This definition of the QC regime reflects the change of the 
structure of the small-$\tilde{\mu}$ expansion. It results in the 
crossover line $\tilde{\mu}\sim e^{-1/(2\delta)}$ between the QC 
and quantum disordered regimes. 
From the other side, the more physical definition of the QC region 
is as a region, where the nonmagnetic contribution to the inverse 
correlation length is larger than the magnetic one. 

 The crossover curves, which correspond 
to the discussed above definitions, are shown in Fig. \ref{figcr-ov}. 
As one crosses the dashed curve in 
increasing $\tilde{\mu}$, the leading nonmagnetic contribution to the 
inverse correlation length becomes $\delta$-independent. 
Below the solid line the magnetic contribution dominates 
(quantum disordered regime), 
whereas above it the correlation length is determined by the 
nonmagnetic contribution.

The form of the distribution $\rho(h)$ and the constant $C_0$ 
correct the position of the solid line, the qualitative behavior being 
unchanged. Rising $\mu$, there are different scenarios of the 
crossover from the quantum disordered to the quantum critical 
regime depending upon which, solid or dashed, line is intersected 
first. If the dashed line is intersected first, then 
one goes from the disordered phase into the region where the leading 
nonmagnetic contribution is already $\delta$-independent, but still 
smaller than the magnetic one.  Further, as the solid line is intersected, 
one enters  the region of the dominance of the nonmagnetic contribution. 
This scenario takes place at comparatively large and at very small $\delta$.   

If the solid line is intersected first, then  one enters the region 
where the nonmagnetic contribution to the correlation length 
dominates, still being $\delta$-dependent. After that, the leading 
contribution to the correlation function becomes $\delta$-independent.

\section{Spin-spin correlation function at small nonzero 
temperature.}

The similarity of finite $\mu$ and 
finite $T$ effects on the quantum phase transition inspires us to 
investigate the question of the interplay of these two factors. 
Thermal fluctuations result in the nonzero occupation of the excited 
states. The fluctuations from the nonmagnetic ground state to the 
magnetic excited state restore, partially, the correlations between 
spins of different magnetic fragments. At the same time, the finite 
temperature destroys the correlations between spins of the 
same magnetic fragment. The competition between these two 
effects determines the physics of the model at small finite 
temperature.

Here the regime $T\ll\mu\ll\Omega_I$ is considered. Then only 
the first excited state may be taken into account. The consideration  
changes in only one point in comparison with the zero-temperature 
case. Namely, in defining the probability of a nonmagnetic site, 
the condition of the nonmagnetic ground state is replaced by the 
probability that the site is in the nonmagnetic state, whether it is 
the ground or the excited state. The probability that the site is in 
the nonmagnetic state can be written as 
\begin{equation}
p(\Omega,\mu,T)=1/(1+\exp[(\Delta-\mu)/T]).
\end{equation}
Here $\Delta-\mu$ denotes the energy gap between the ground 
and the first excited state. $\Delta=\Omega$ if $\Omega$ is a 
transverse field, $\Delta=\Omega+J_{R,L}$ if $\Omega$ is a 
bond, and $J_{R,L}<\Omega$ is the other bond adjacent to 
the site.  The finite temperature therefore causes the smearing 
of the probability of the  nonmagnetic  state $\vartheta(\mu-\Omega)$ 
over the region of order $T$ around $\mu=\Omega$. 

The formulas for the probabilities $w_J$ and $w_h$, 
Eqs. (\ref{wj}) and (\ref{wh}), become 
\begin{equation}
w_J=2\int_0^{\Gamma_h}d\Gamma 
\int_{-\ln\left(\mu/\Omega_I-e^{-\Gamma}\right)}^\infty d\xi 
S(0,\Gamma) S(\xi,\Gamma)/ 
(1+\exp[(e^{-\Gamma}+e^{-\zeta}-\tilde{\mu})/\tilde{T}]) , 
\end{equation}\label{wjT}
\begin{equation}
w_h=\left[\int_0^\infty d\zeta S(\zeta,\Gamma_h)\right]^2  
(1+\exp[(h-\mu)/T])^{-1}, 
\end{equation}\label{whT}
where $\tilde{T}\equiv T/\Omega_I$.

The temperature behavior of the correlation length was analyzed 
numerically. The dependence of $W_h(T)$ on temperature 
is monotonously decreasing at small transverse field $h$, but 
it changes to the nonmonotonous at larger fields $h\sim\mu$. 
The resulting temperature dependence of the probability of 
the nonmagnetic state $W(T)$ depends strongly upon the 
original distribution of transverse fields and can also be  
nonmonotonous. 

The effect of  breaking of the magnetic correlations by thermal  
fluctuations is particularly strong in the quantum disordered phase 
and at the critical point, where it contributes to the inverse 
correlation length as $\pi^2/(\ln\tilde{T})^2$ and 
$\pi^2/(2\ln\tilde{T})^2$, respectively, (see I). 
This effect  dominates in the quantum disordered 
phase and at the critical point, and leads to a monotonously decreasing 
overall temperature dependence of the correlation length. 

In the ordered phase, however, the temperature breaking of the 
correlations is weaker and contributes to the inverse correlation 
length as $4\delta^2 (\tilde{T})^{2|\delta|}$ (Ref.I). In this 
phase the temperature dependence of the inverse correlation 
length is nonmonotonous if the distance from the critical point 
$\delta$ exceeds some finite value (see Fig. \ref{figxto}).

\section{Dependence of averaged filling on chemical potential 
$\nu(\mu)$. One-fermionic density of states.}

The obtained expressions for the probability of the 
nonmagnetic site allow us to calculate the dependence of 
the average filling on $\mu$. For $\mu>0, T=0$, each 
nonmagnetic site is doubly occupied and the average 
filling can be written as  
\begin{equation}
\bar{\nu}(\mu)= 
lim_{l\rightarrow\infty}(1/l)\sum_{i=1}^l(1-W+2W)=1+W(\mu),
\end{equation}\label{nu}
where $W$ is the probability of a nonmagnetic site ($\nu=2$) 
and $1-W$ is the probability of a magnetic site ($\nu=1$). 
Substituting the expressions for $W(\mu)$,  we obtain for 
the average filling the following formula:
\begin{equation}
\bar{\nu}\approx 1+ B(\mu, b)\frac{e^{2\Gamma_\mu\delta} 
\sinh^2(C_0\delta)}{\sinh^2[(C_0+\Gamma_\mu)\delta]},
\label{numu}\end{equation} 
where $B(\mu, b)=\int_0^\mu \rho(h) dh$ is a specific for 
the original distribution of the transverse fields factor, 
\begin{equation}
B(\mu,b)=\left\{
\begin{array}{cc}
\mu/b & \  \mbox{for the distribution (a)} \\
1-e^{-\mu/b} & \ \mbox{ for the distribution (b)}.
\end{array}
\right.
\end{equation}
Equation (\ref{numu}) describes the average filling in the 
quantum disordered phase at $\delta>0$ and in the ordered phase 
at $\delta<0$. In the limit $\delta\rightarrow 0$, one obtains 
$\bar{\nu}(\mu)$ at the critical point as
\begin{equation}
\bar{\nu}\approx B(\mu, b)\frac{C_0^2}{(C_0-\ln\tilde{\mu})^2}.
\label{numuc}\end{equation}

Taking the derivative $d\bar{\nu}/d\mu$, one can analyze the 
behavior of the average density of states. Close to the half-filling 
($\mu\ll 1$), one obtains  the average density of states in the form 
\begin{equation}
\frac{d\bar{\nu}}{d\mu} = (1/b) 
\frac{\tilde{\mu}^{-2\delta}\sinh^2(C_0\delta)}{\sinh^2[(C_0-
\ln\tilde{\mu})\delta]} \left(1+\frac{2\delta \tilde{\mu}^\delta 
e^{-C_0\delta}}{\sinh[(C_0-\ln\tilde{\mu})\delta]} \right)
\end{equation}
for both distributions  (a) and (b) of transverse fields. 

In the disordered phase ($\delta>0$), the density of states at 
half filling is finite and equals 
\begin{equation}
\frac{d\bar{\nu}}{d\mu}= (4/b) 
e^{-2C_0\delta}\sinh^2(C_0\delta).  \end{equation} 
The behavior of the density of states in the quantum disordered 
phase is highly nonuniversal and depends strongly on the original 
distribution of transverse fields. For example, the average density 
of states would acquire a gap if there were a gap in the original 
distribution of transverse fields, i.e., a nonzero minimal value of the 
original transverse field.

In the ordered phase ($\delta<0$), the behavior of the density of 
states is more universal in the sense that there is a gap in the 
density of states at half filling if the function $B(\mu)$ grows 
more slowly than $\mu^{-4|\delta|}$ as $\mu$ goes to zero. The 
density of states for the distributions (a) and (b) reads 
\begin{equation}
\frac{d\bar{\nu}}{d\mu}= \frac{4}{b}\sinh^2(C_0\delta)
e^{-2C_0|\delta|}(1+4|\delta|)\tilde{\mu}^{4|\delta|}.
\end{equation}

The soft gap behavior of the average density of states at half 
filling survives at the critical point as well. Here the condition 
for the function $B(\mu)$ is that it should grow more slowly 
than $(\ln\mu)^2$ as $\mu$ goes to zero. The expression for the 
average density of states for cases (a) and (b) at the critical point 
looks like 
\begin{equation}
\frac{d\bar{\nu}}{d\mu}= (1/b)\frac{C_0^2}{(C_0-\ln\tilde{\mu})^2} 
\left(1+\frac{2}{C_0-\ln{\mu}}\right).  \end{equation}

\section{Conclusion}

We showed that the extension of the RTFISC model on the 
fermionic space, introducing statistical charge fluctuations, 
brings about interesting features in the magnetic as well as in the 
fermionic characteristics. 
Strong  mutual effects of magnetic and charge fluctuations show 
up in  the power-law scaling of the contribution of charge fluctuations 
to the inverse spin-spin correlation length 
with the distance from the magnetic quantum critical point and 
also in the appearance of the quantum critical region in the phase 
diagram.

Although the characteristics in the charge sector (percolating 
correlation length, fermion density of states) behave nonuniversally, 
they contain a universal factor, which can be separated from the 
nonuniversal one. The nonuniversal features enter through the 
original distributions of transverse fields, whereas the bonds are 
described by universal scaling distributions. Then, it is natural that 
the nonuniversal behavior is particularly strong in the quantum 
disordered phase, where the  transverse fields dominate. 

The results on the one-fermion density of states suggest different 
structures of electronic states in the ordered and in the quantum 
disordered phases in the model with electron hopping. The investigation 
of this model is left for the future. \\

The author acknowledges support by the Deutsche Forschungsgemeinschaft 
through the Sonderforschungsbereich 410.

$\ast$ Present address: Department of Physics and Astronomy, 
Ohio University, Athens, OH 45701.

\newpage
\begin{figure}
\epsfig{file=cr-ox-d.ps,width=10cm,bbllx=-20pt,bblly=390pt,bburx=270pt,bbury=748pt,angle=0}
\vspace{.5cm}
\caption{Crossover lines $\tilde{\mu}(\delta)$ between different 
regimes in the phase diagram calculated for the distribution (b) with 
the width $b=1/2$. Above the solid line the nonmagnetic contribution 
to the inverse correlation length dominates. Above the dashed line the 
leading nonmagnetic contribution is $\delta$-independent. $C_0=1$. 
\label{figcr-ov}} 
\end{figure}

\newpage
\epsfig{file=xtobe-d.ps,width=10cm,bbllx=-20pt,bblly=390pt,bburx=270pt,bbury=748pt,angle=0}
\vspace{.5cm}
\begin{figure}
\caption{The temperature dependence of the correlation  length in the ordered phase 
becomes nonmonotonous as the distance $|\delta|$ exceeds some value. $\xi(T)$ for the 
distribution 
(b) with $b=1$. $C_0=1$, $\delta=-0.4$, $\tilde{\mu}=0.3$.
\label{figxto}}
\end{figure}

\end{document}